# Piezonuclear reactions – do they really exist?

G. Ericsson[1], S. Pomp[1,*], H. Sjöstrand[1], E. Traneus[2]

[1]Department of Physics and Astronomy, Uppsala University, Uppsala, Sweden

[2]RaySearch Laboratories, Sweden

[*]Corresponding author. *E-mail address*: Stephan.Pomp@physics.uu.se (S.Pomp)

**Abstract**

In a number of recent articles in this journal F. Cardone and collaborators have claimed the observation of several striking nuclear phenomena which they attribute to "piezonuclear reactions". One such claim [Phys. Lett. A 373 (2009) 1956] is that subjecting a solution of $^{228}$Th to cavitation leads to a "transformation" of thorium nuclei that is $10^4$ times faster than the normal nuclear decay for this isotope. In a "Comment" [Phys. Lett. A 373 (2009) 3795] to the thorium work, we have criticized the evidence provided for this claim. In a "Reply" [Phys. Lett. A 373 (2009) 3797] Cardone et al. answer only some minor points but avoid addressing the real issue. The information provided in their Reply displays a worrying lack of control of their experimental situation and the data they put forward as evidence for their claims. We point out several shortcomings and errors in the described experimental preparations, set-up and reporting, as well as in the data analysis. We conclude that the evidence presented by Cardone et al. is insufficient to justify their claims of accelerated thorium decay (by "piezonuclear reactions" or otherwise). We also briefly discuss the role of the physics community (peer review) in the evaluation of new discoveries and claims.

## 1 Introduction

In a number of recent publications in Physical Letters A [1, 2, 3], F.Cardone and collaborators have reported on experimental results which they claim show evidence for a new type of nuclear reaction (decay), namely, "piezonuclear reactions". We have scrutinized one of these publications [1] in some detail and found serious shortcomings in the experimental procedures, data analysis and reporting of the work. In their paper, Cardone et al. make a very specific, and quite extraordinary claim about the laws of nature, namely, that the decay rate (or "transformation", as the authors sometimes prefer to call it) of the isotope $^{228}$Th in a water solution can be increased by a factor of 10000 compared to its natural decay rate by exposing it to cavitation, i.e., sound waves at 20 kHz and 100W. Since this is the first time, to our knowledge, that such a claim is presented in a widespread physics journal we find it important that it receives careful attention by the physics community, and this for two reasons. Firstly, if true this new "piezonuclear" phenomenon would have far-reaching implications for our understanding of nature in general and nuclear physics in particular. The practical applications could be immense in, e.g., the fields of nuclear energy and regarding the treatment of spent nuclear fuel, just to mention a few obvious examples. Secondly, the issue touches on important aspects of what is the "scientific method", both regarding the responsibilities of proponents of a new theory or discovery to present their work in the clearest and most accessible way, and also the role of the physics community in the peer review of such claims both before and after publication.

## 2 Overview of the experiment and our criticism

In a previous Comment [4] we outlined some of the findings from our investigation of the material presented in [1] and put forward a few suggestions to improve the work. In a Reply [5] to our Comment, Cardone et al. address some but not the most important points raised by us. In fact, instead of clarifying the situation, the Reply by Cardone et al. raises further, very severe questions about the experimental work and data analysis methods used.

We first recall the basic facts of the matter. Cardone et al. claim that cavitation of a thorium solution speeds-up its "transformation" by a factor $10^4$ compared to its natural decay rate. As evidence for this extraordinary claim they present two sets of experimental measurements. The first is a set of twelve (12) CR39 detectors exposed to thorium solutions. Four of these detectors measured non-cavitated reference samples and eight measured samples during cavitation. As corroborating evidence, thorium concentrations were measured for 3 (of the 4) non-cavitated solutions, and for 3 (of the 8) cavitated solutions; these measurement were done with accelerator mass spectrometry after the cavitation exposure. In other words, in order to substantiate a claim regarding a new major decay mode of thorium, Cardone et al. present pictures of 12 CR39 detectors and post-cavitation concentration measurements of 6 thorium solutions.

In this paper we will first summarize our four main concerns. We then discuss these points in some more detail after which we make some general remarks concerning the paper and Reply by Cardone et al. and come to a conclusion.

Our four main concerns are the following:

1) Statistical analysis and error propagation: The original paper does not contain *any* discussion of the uncertainties in the final values that are put forward as support for the claim. Neither does it contain *any* statistical analysis of the experimental data that could serve as a test whether the null hypothesis (i.e., that cavitated and non-cavitated samples are drawn from the same distribution) can be rejected or not. In fact, Cardone et al. discuss no alternative explanation for their result whatsoever. We show in Ref. [4] and below that, both regarding the CR39 and the concentration data, the null hypotheses cannot be rejected.
2) The experimental method: The experimental conditions of the reported work are not well described, neither in Ref. [1] nor in [5], and we have partly to rely on guesses on what might have been done. From what is presented it seems to us that the experiment was not adequately planned and prepared. For example, considering their limited sensitivity in the present application, the use of CR39 detectors seems far from optimal. No indication is given if the CR39 counting statistics was estimated before the experiment and if the obtained data is reasonable with the present experimental set-up. The known difficulty in preparing thorium solutions of uniform concentration should have been taken into account when preparing the experiment; thorium solutions of higher concentration could have been used in order to avoid this problem. Exposure times of the CR39 detectors were apparently too short to give adequate counting statistics; exposure times could have been increased considerably compared to the 90 minutes now used. Complementary measurements of gammas should have been undertaken. Finally, variations of the main experimental parameters should have been tested to investigate the alleged phenomenon and sources of uncertainty.
3) Data selection and presentation: The extraordinary claim made by Cardone et al. is backed up by a very limited set of experimental data, both regarding the CR detectors and the concentration measurements. The experiment should have been conducted in a way to ensure sufficient statistics and sensitivity. Furthermore, only a sub-set of the available data is presented. For example, of the "12 identical samples" (having, in fact, concentrations ranging by up to a factor of three) only 6 post-cavitation concentration values are presented, even though it is stated that a more complete data set exists. In order to draw any conclusion from the concentration measurements it is crucial to present the concentrations of *all* samples, both *before* and *after* cavitation.
4) Background measurements: Probably our most severe concern is with respect to the data Cardone et al. present in Fig. 1 in their Reply. This picture of two new CR39 plates is supposed to show background measurements with pure bi-distilled water; (a) inside the vessel and (b) outside the vessel. We note that (a) does not resemble the pictures in Fig. 1 of Ref. [1] in any way. We therefore fail to see how this can constitute a valid background measurement (in the normal sense of that term) for the CR39 detectors presented in the original paper. The case of Fig. 1b in [5] is even more disturbing since it is an *exact copy* of a CR39 picture that Cardone et al. presented already in their original paper, then claimed to be one of the CR39 exposed to cavitation in a thorium solution (Fig. 1 [1]; cavitated samples, second group, second frame from above). Obviously, a single CR39 plate cannot at the same time constitute a signal *and* a pure background measurement!

This short summary should make it clear that there are severe shortcomings in the experimental and data analysis methods used by Cardone et al. However, in order to avoid misunderstandings, we will in what follows elaborate on each of the four points above and make some general comments and observations about the original paper and the Reply by Cardone et al.

## 3 Detailed criticisms

### 3.1 Statistical analysis and error propagation

We have shown in our Comment that based on the provided data, the null hypothesis, namely that there is no difference in the decay rate of thorium in cavitated and non-cavitated samples, cannot be rejected. A t-test based on the data in Fig.1 [1] gives a p-value of 0.26, meaning that under the assumption that the two series of CR39 samples are drawn from the same distribution, in about one quarter of such experiments one would expect to obtain a distribution of samples as extreme or more extreme as the one observed here. Since a much lower p-value is required (we have suggested 0.01) in order to reject the null hypothesis, our conclusion is that it *cannot* be rejected. Cardone et al. try to soften the requirement on the p-value and suggest accepting values as high as 0.05. In their own calculation, which in fact we cannot reproduce[1], they obtain a p-value of 0.065 which, however, *also* fails their own criterion. The reasonable thing to do at this stage would be to admit that the strength of the evidence is not sufficient by the set standard (arbitrary set by themselves *after* our Comment), and that additional experimental data are required before the work can be presented and any conclusions drawn. Instead, Cardone et al. are content to reach a value "very near to the 5% usually accepted by the scientific community" and even argue that their "statistical result […] has actually a great statistical meaning" [5]. To us, such statements are just smoke screens to hide the fact that their data do not give any reason to reject the null hypothesis.

---

[1] In the Reply of Cardone et al. it is claimed that, using a two-tailed t-test within the software environment "R" [6], a p-value of 0.065 is obtained from the alpha track (CR39) data presented in Figure 1 [1]. Trying to reproduce this result we performed a two-tailed t-test on the number of counts for the two measurement series: x=[1 0 1 1] and y=[1 0 1 0  0 0 0 1] using the R-command: `t.test(x,y,alternative=c("two.sided"),var.equal=TRUE)`. We obtain a p-value of 0.26, the same value as we reported in our Comment [4], then performing the t-test using the Matlab function "`ttest2()`". Using a one sided t-test with the alternative hypothesis that the cavitated sample has values lower than the non-cavitated sample (using the R-command: `t.test(y,x,alternative=c("less"),var.equal=TRUE)`) we obtained a p-value of 0.13. The same result was also obtained using Matlab.

A similar case can be made based on the concentration values in Tables 1 and 2 of Ref. [1]. Our calculation of the ratio, R, of concentration values for non-cavitated and cavitated samples gives $R = 2.1 \pm 2.6$ (or $2.1 \pm 1.5$ depending on the assumption we have to make regarding the underlying uncertainties) is not challenged in the Reply by Cardone et al; in fact, the issue isn't even mentioned in their Reply. Again it is clear that the null hypotheses, i.e., in this case $R = 1$, cannot be rejected.

As is clear from the above, statistical tests applied to the presented data give no justification to reject the null hypothesis on any relevant significance level. Let us add that we are fully aware that the t-test might not be the optimal tool under the present circumstances. However, what should be perfectly clear is that the scarce experimental evidence presented by Cardone et al. is not sufficient to safely reject the null hypothesis. We are truly concerned that Cardone et al. fail to acknowledge this fact.

### 3.2 Experimental method

The experimental methods and procedures chosen by Cardone at al. seem far from optimal to study the possibility of accelerated thorium "transformation" from cavitation. In many respects, what is presented looks more like a preparatory laboratory test than a serious nuclear physics experiment.

The use of CR39 detectors as the main radiation measurement tool means that only a very limited volume (which might not be representative) of the thorium solutions is actually probed for alpha decays. We note that the range of 5.5 MeV alpha particles in water is about 40 µm. The CR39 measurements should have been complemented by other measurements of, for example, gammas. Furthermore, the exposure time of the CR39 detectors is only 90 minutes, but there seems to be no compelling reason why measurements should be done only during cavitation and couldn't be extended for much longer times. Since scarcity of data is indeed a problem in the presented work, measurements of both cavitated and non-cavitated solutions should have been extended much beyond the present 90 minutes to give CR39 data of sufficient statistical significance. In that respect, the present measurement could have served as an indication of the collection times needed to obtain a more conclusive result.

Measurements of the thorium concentrations were done only after cavitation. With the known limitation in preparing samples of uniform concentration, concentration measurements by AMS should have been done and presented for all samples both *before and after* cavitation.

The known limitation in reproducibility of the concentrations (estimated by Cardone et al. to be ± 0.01 ppb) of the samples should have been taken into account when planning the experiment. If higher thorium concentrations had been used, this limitation could have been circumvented or at least reduced.

The criteria for identification of alpha tracks in the CR39 detectors should be clearly presented. In response to our question on this issue, Cardone et al. refer to their colleague (and co-author of Ref. [3]) Dr. G. Cherubini "who prides himself on decades long experience in this field". We are of course happy that Dr. Cherubini finds pride in his work and has had a long career. However, these are not valid arguments in a scientific discussion; we cannot resort to solely invoking the "long experience", expertise or "pride" of our colleagues when asked for clarifications on experimental techniques. We have to base our presentations on clear descriptions of procedures and solid evidence. For some reason Cardone et al. are silent on the question of providing further understanding of the procedures for identifying alpha particles in the CR39 detectors. The new CR39 data they present in Fig. 1 of their Reply [5] only adds further confusion, as we discuss below.

### 3.3 Data selection and presentation

The fact that only a limited set of data is presented is another major problem. Cardone et al. write in [1] that "12 identical samples" have been prepared but specify later that the concentrations in fact range from 0.01 to 0.03 ppb, i.e., they differ by up to a factor of three. Nowhere do they list the concentrations of the 12 samples *before* cavitation even though this information is crucial for judging the results. Furthermore, they show concentrations and results from AMS measurements only for 6 out of the 12 samples, although apparently a more complete data set exists. The data points that are presented *exhibit a large spread* even within each of the groups (cavitated and non-cavitated). This large variation in the concentrations is clear indication that the result, i.e., the ratio, depends very much on the selection of samples and on the initial concentrations. We cannot refrain from wondering if Cardone et al. cherry pick data that give the answer they have already set out to find.

### 3.4 Background measurements

Cardone et al. state in their Reply that background measurements with "mere bi-distilled deionised water" were in fact performed. They also present two new pictures of CR39 detectors allegedly exposed to cavitation under these conditions; one detector was placed "on the bottom of the vessel" (Fig. 1a [5]) and the other (Fig. 1b [5]) "below the vessel". Cardone et al. further state that "no trace on them [i.e., the CR39 detectors in their Fig. 1, our comment] does vaguely resembles those observed in the thorium experiment [1]". This statement poses two big problems for us.

First, since Fig 1a, allegedly of a CR39 placed inside the vessel filled with (only) pure water and exposed to cavitation, does indeed not resemble any of the CR39 data presented in the original paper [1], we fail to see how this can constitute a valid background measurement for the CR39 detectors presented in that paper. The normal use of the word “background” in such circumstances would imply that any signal should be superimposed “on top of” this background. However, none of the CR39 plates presented in [1] has a structure even remotely similar to this “background measurement”. Instead of contributing to clarifying the situation, Cardone et al. add even more confusion to the issue.

Secondly, and this is even more serious, a closer examination of Fig. 1b in their Reply [5] reveals that it is *exactly the same* CR39 picture as they already presented in their original paper [1] as one of the CR39 exposed to cavitation in a thorium solution (Fig. 1: cavitated samples, second group, second frame from above). So, not only does this CR39 picture *closely resemble* the data presented in the original paper (in contradiction to their previous statement), but it is in fact *an exact copy*! Obviously, a single specific CR39 picture cannot both represent a background measurement taken outside of the cavitation vessel filled with pure water *and* a measurement inside the vessel during cavitation of a thorium solution.

Without proper recordkeeping, careful scrutiny of published data and clear written statements, a scientific discussion cannot be undertaken and an exchange of the type we are trying to conduct here is meaningless. Possibly, Cardone et al. will dismiss this serious methodological flaw as a printing mistake or the like. However, we note that exactly the same figures are used in Ref. [7] (as indeed also stated by the Authors) so they have on two occasions examined these pictures and concluded they are valid evidence in their publications. In our minds this casts serious doubts on the whole set of experimental data presented by Cardone et al. in their original paper [1]. Under what conditions were the CR39 detectors really used? Is there proper record keeping of all the CR39 plates and other experimental procedures? Which of the other CR39 plates in the original paper were exposed to only pure water? With such uncertainties and mistakes in the experimental material presented by Cardone et al., further discussions on this matter are pointless.

**4 General remarks on the paper**

The methodological problems we have discussed above are further highlighted by what Cardone et al. write under section 2.4 in their Reply: “it was not our aim to carry out a precise inferential statistical analysis”. What then is the point of performing an experiment and submitting a paper for publication if one does not carry out a proper analysis of the data obtained? They continue in the same section: “Actually, as already stressed, our paper must be inserted in the context of the evidence from other analogous experiments, in which cavitation was shown to induce 'anomalous' nuclear effects." To us, this indicates that Cardone et al. take the existence of these "anomalous" nuclear effects for granted and deem it unnecessary to test their hypothesis.

Instead of seriously considering ways to improve their work, Cardone et al. dismiss our suggestion, for example, to measure the gamma radiation emitted by the thorium solutions under study as "invalidated by the fact that piezonuclear reactions occur without gamma emission". To us there are two mistakes in this statement. First, and let us be very clear on this, these "gamma-less piezonuclear reactions" are at this point hypothetical and by no means an established physical phenomenon. Cardone at al. are making a remarkable claim about the physical reality, namely, that under cavitation, thorium "transforms" at a much higher rate than its natural decay rate, and that this "transformation" occurs without gamma radiation. This specific claim about the physical reality has to be verified in its own right with clear experimental evidence presented by those making the claim. There exist no accepted "facts about piezonuclear reactions" that can diminish this requirement. Secondly, if thorium is really removed from the solution by "piezonuclear transformations" it should be possible to establish a difference in the emission rate of characteristic gammas (from the normal thorium decay chain) between cavitated and non-cavitated samples. Such gamma measurements would be of great additional value in this type of investigation, and could also shed light on the temporal correlation between cavitation and thorium "transformation", if indeed there is any.

In this context we must also note that while Cardone et al. are quite vague about the nature of the thorium "transformation" in the original paper ("It is still open the question whether the effect […] was simply to accelerate its natural decay process, or it underwent other types of transformations (like fission process).") they now present a quite different story where "thorium is halved [sic!] by means of piezonuclear reactions without increasing its radioactivity." We fail to see what added evidence has been presented in their Reply that can motivate such a different, much more specific, conclusion.

Cardone et al. criticize us for being ignorant about the wider research stream of piezonuclear reactions. However, it should come as no surprise to Cardone et al. that far-reaching claims, based on highly speculative new physics, are viewed with a great deal of scepticism by the general physics community. It is normally the duty of the proponents of a new physics phenomenon to provide irrefutable evidence for their claim. And it is the role of the general scientific community to be sceptical in such matters, to force the proponents to acquire the required evidence and present their case in the clearest and most convincing way. This position is in general not based on prejudice or conspiracy, but is a simple consequence of the position that "extraordinary claims should be matched with extraordinary evidence", something that proponents of phenomena like cold fusion, sono-fusion, and, in this case, “piezonuclear nuclear transformations” have so far been unable to provide. It therefore does little to convince the community when Cardone et al. cite their own work at length (including a paper flagged as “question mark paper” like Ref. [6] in their original paper), or papers on other highly speculative topics. And those references do nothing to support the claim made in the paper under discussion here, which is a very specific claim about the physical reality regarding thorium decay, supposedly based on presented experimental data.

## 5 Discussion and Conclusions

In this paper we show that the claim by Cardone et al. of accelerated (and since the Reply even gamma-less) ”transformation” of thorium due to cavitation is not substantiated by the experimental evidence presented. We have also shown that the experimental procedures and the treatment of data are below the standards normally accepted by the physics community. Furthermore, we have even discovered serious errors and shortcomings in the presented data. Under these circumstances we must conclude that the claims made in papers [1] and [5] should be considered as mere speculation on the part of the authors.

Peer review of submitted scientific papers is commonly regarded as one of the cornerstones of the scientific process. Although many problems are associated with this system and there is a continuous discussion within the scientific community regarding its benefits, problems and possible improvements [8], peer review in one form or another is still considered an essential ingredient in the evaluation of new discoveries and claims. The most common form of this activity is the system of anonymous referees solicited by the editor of the scientific journal to which the paper is submitted. Such refereeing can normally weed out submissions which are below the accepted standard for scientific publishing. However, we would also like to emphasize the responsibility we all have as members of the scientific community, to actively scrutinize and comment on material published in the fields of our individual expertise. As a matter of fact, some students of the “scientific method” see such public scrutiny and debate as one of the possible bases for defining what constitutes science [9]. With the rapid growth in number of publications and the increased workload on scientists, the possibility to find referees of sufficient expertise and with sufficient time at the appropriate moment is becoming an increasing problem. This situation can also result in an attitude of the community to ignore publications that are considered as sub-standard simply because there is not enough time to undertake the work of close examination and criticism. However, in a time when science is questioned and misused on different levels, we will make ourselves a disfavor if we do not require all participants in the scientific endeavor to adhere to the accepted high standards of the field. The task we have undertaken here should be seen in this context and, we hope, inspire others to take an active part in the scientific discussion.

**Acknowledgements**

We thank Prof. Leif Nilsson and the colleagues at the Division of Applied Nuclear Physics, Uppsala University, for inspiring and enlightening discussions on this issue.